\documentclass{PoS}

\newcommand{\obar}[1]{\overline{#1}}
\newcommand{\fm}{\mbox{\,fm}}
\renewcommand{\O}{\mathcal{O}}

\newcommand{\ve}[1]{\mathbf{#1}}

\renewcommand{\>}{\rangle}
\newcommand{\<}{\langle}

\PoS{PoS(LAT2005)217}

\title{NRQCD results on the MILC extra coarse ensemble}
\ShortTitle{NRQCD results on the MILC extra coarse ensemble}

\author{\speaker{Ian Allison}, Christine Davies\\

  Department of Physics and Astronomy \\
  University of Glasgow,\\
  G12, 8QQ \\
  UK

  E-mail: \email{i.allison@physics.gla.ac.uk}
}
\author{Alan Gray\\

  Edinburgh Parallel Computing Center \\
  University of Edinburgh, \\
  EH9, 3JZ \\
  UK 
}

\abstract{We present preliminary results using NRQCD to describe heavy
  quarks on the MILC $2+1$ flavour dynamical \emph{extra coarse}
  ensemble. We calculate the spectra of low lying states in
  bottomonium to complement earlier results on the finer MILC
  ensembles. We then exploit the coarseness of the lattices to
  calculate charm propagators using NRQCD. These are used to examine
  the charmonium spectrum and to calclate the mass of the $B_c$ using
  NRQCD. Finally we look breifly at the $B_d$ and $B_s$ systems using
  the imporoved staggered formalism to describe the light valence
  quarks.}

\FullConference{XXXIIIrd International Symposium on Lattice Field Theory\\
  25th-30th July 2005 \\
  School of Mathematics, Trinity College, Dublin, Ireland
}

\begin{document}
\section{Introduction} 
\label{sec:introduction}
Heavy quark physics presents the lattice with an invaluable opportunity
to provide both theoretical insight and the key to some precise
results from experiment. The use of effective field theory techniques
such as NRQCD~\cite{Lepage:1992tx} allow heavy quarks to be described
on the lattice avoiding the large discretisation errors of
relativistic formulations.  

The basic idea of these techniques is that simulations should be
performed on lattices coarse enough that the cutoff $\pi/a$ separates
the heavy quark mass\footnote{Throughout this work, the term heavy
  quark refers to either the $b$ or $c$ quark.} scale $M$ from the
dynamical scales $Mv$, $Mv^2$, where $v$ is the heavy quark velocity
inside the bound state. This allows a non-relativistic formulation of
QCD to simulate only the dynamical scales important to heavy mesons,
with the physics at the heavy quark mass scale (and above) being
encoded into matching coefficients.

The other key component of the calculations presented here is the use
of $2+1$ flavour dynamical configurations provided by the MILC
collaboration~\cite{Bernard:2001av}. These configurations employ the
$a^2$ improved ``AsqTad'' improved staggered quark action and a highly
improved gluon action to describe the vacuum. Our calculations look at
the coarsest ensemble generated which is comprised of around $600$
$16^3\times 48$ lattices with spacing $a\sim 0.17\fm$, intended to
complement the coarse $(a\sim 0.12\fm)$ and fine $(a\sim 0.09\fm)$
ensembles on which the majority of results have been extracted.  Part
of the motivation for our calculations was to determine the
feasibility of the extra-coarse ensemble for extraction of physical
results in the presence of possible large discretisation errors.

In section~\ref{sec:form-techn-deta} we outline the formulation used
for our calculations, then in section~\ref{sec:upsilon-psi-heavy} we
look at the $\Upsilon$ and $\psi$ spectra. In
section~\ref{sec:b_c-mass} we calculate the mass of the $B_c$ meson
using NRQCD for both the $b$ and $c$ quarks. Finally, in
section~\ref{sec:heavy-light-results} we look at the $B_s$ and $B_d$
mesons by combining an NRQCD $b$ quark with an improved staggered
light quark.

\section{Formulation and Technical Details}
\label{sec:form-techn-deta}

A potential model calculation using a logarithmic
potential~\cite{Quigg:1979vr} can be used to extract the typical
velocity of the heavy quarks in their bound states. Doing so gives
$v_b^2\sim 0.1$ for $b\obar{b}$, $v_c^2 = 0.25$ for $c\obar{c}$ and
$v_b^2\sim 0.04, v_c^2\sim 0.4$ for the $B_c$ meson. These values
justify a non-relativistic description, with the possible exception of
the $c$ quark of the $B_c$ system. This work makes use of the non
relativistic formulation of QCD (NRQCD) developed
in~\cite{Lepage:1992tx}. In this formulation, the operators of the
heavy quark hamiltonion are classified according to a power counting
in the velocity of the heavy quark in the bound states; $v$.  We use
the action given in\cite{Lepage:1992tx} which is correct through
$\O(v^4)$.

We fit the mesonic corellators using the constrained curve fitting
method of~\cite{Lepage:2001ym}. This method employs Bayesian
techniques which utilise \emph{prior} information to guide the fit,
allowing \emph{all} of the excited states present in the initial few
time slices of correlator to be fitted. This improves the statistics
of the fits while reducing the sensitivity to the number of
exponentials used.

\section{The $\Upsilon$ and $\psi$ heavy quarkonium systems}
\label{sec:upsilon-psi-heavy}
The majority of calculations using NRQCD describe $b$ quark systems,
see for example~\cite{Gray:2005ur}. In the quantity $aM_Q$, $M_Q$ for
the $b\obar{b}$ is a relatively large number ($M_Q\gg 1$), so that
even with present day lattices $aM_q$ is quite large. On the same
lattices, the $c\obar{c}$ system requires an $M_Q$ of less that $1.0$
and inverse powers of $M_Q$ in the NRQCD evolution equation then cause
numerical instability, preventing results from being extracted. The
advantage of the extra coarse configurations is that $M_Q=1.0$
produces mesons whose kinetic mass is close to that of the $J/\psi$
particle while the evolution equation remains stable.

\begin{figure}[hb]
  \begin{minipage}[l]{0.49\linewidth}
    \begin{center}
      \includegraphics[width=1.0\linewidth]{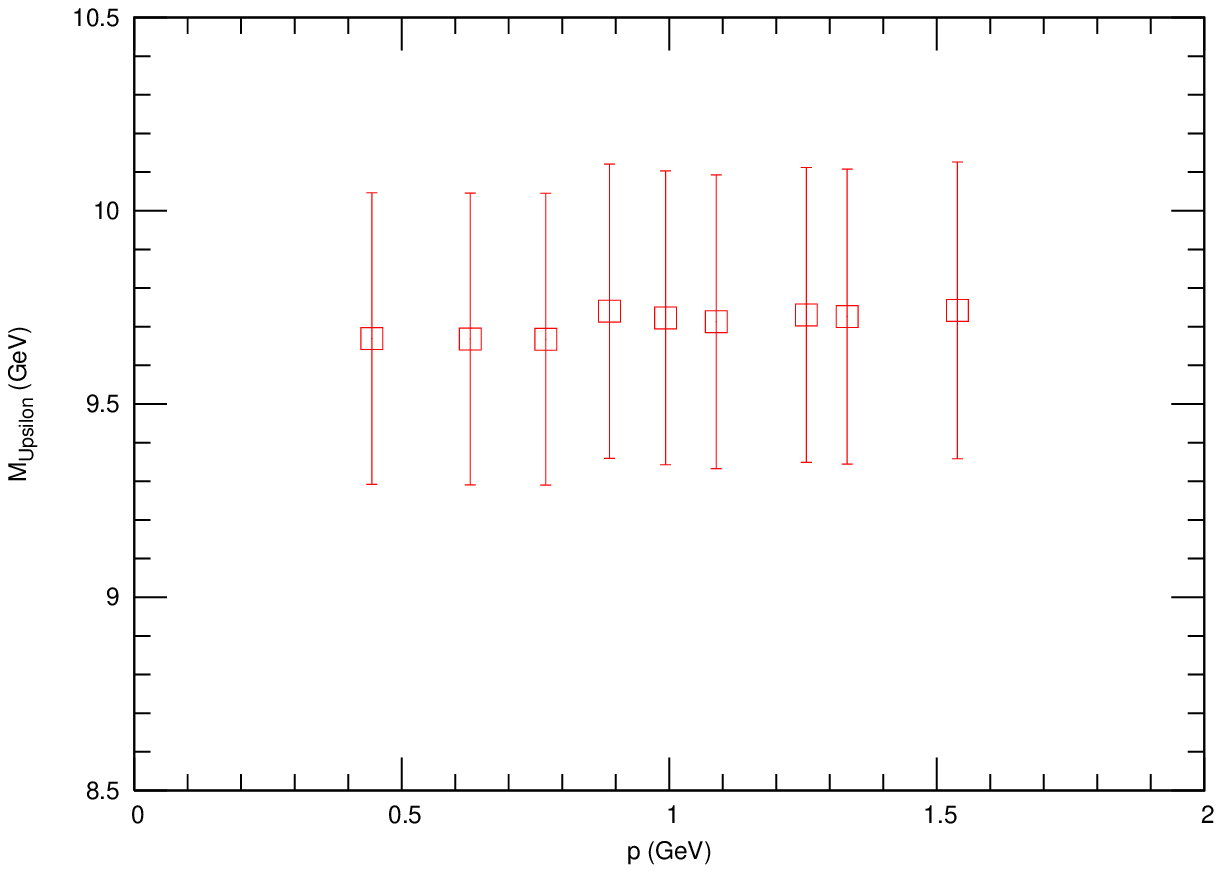}
      \caption{Kinetic Mass of the $\Upsilon$}
      \label{fig:kinups}
    \end{center}
  \end{minipage}
  \begin{minipage}[r]{0.49\linewidth}
    \begin{center}
      \includegraphics[width=1.0\linewidth]{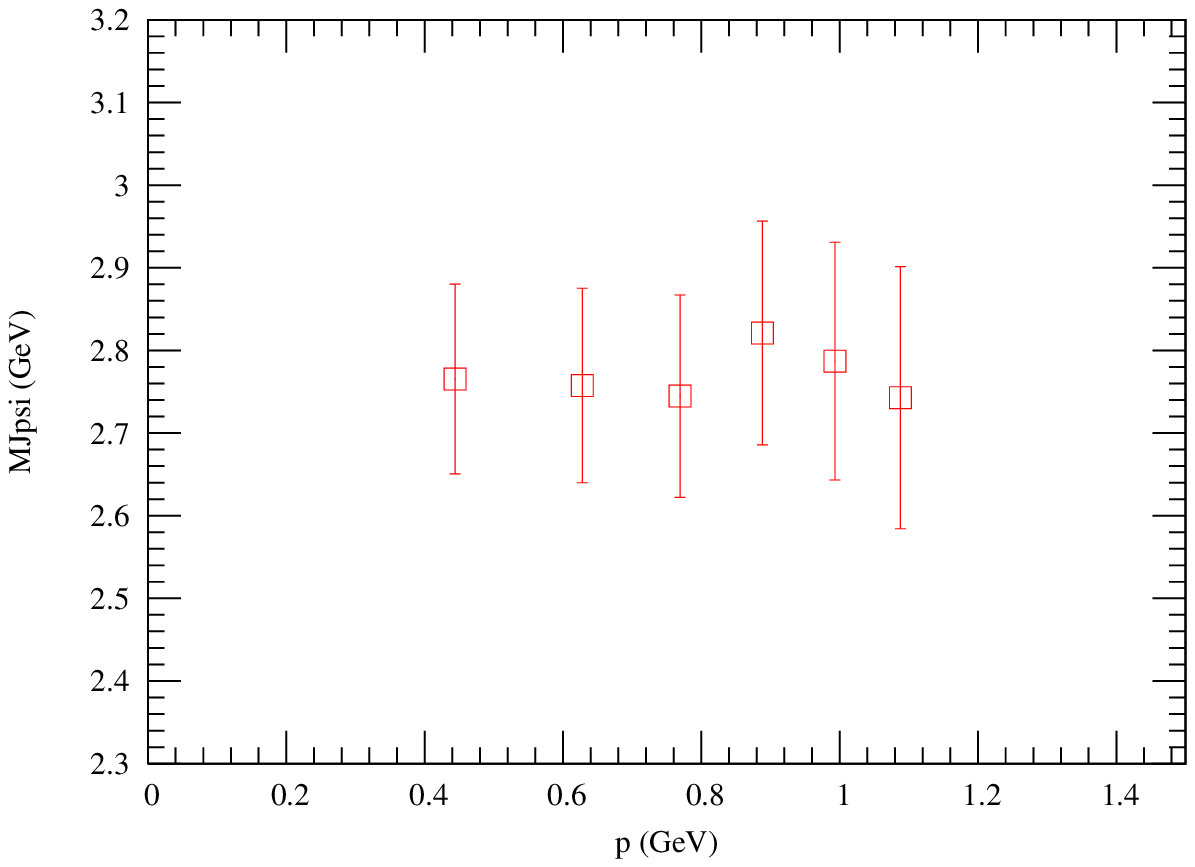}
      \caption{Kinetic Mass of the $J/\psi$}
      \label{fig:kinchm}
    \end{center}
  \end{minipage}
\end{figure}

In NRQCD, the zero of energy is shifted, so that the energy of a meson
with momentum $\ve{p}$ is 
\begin{equation}
  \label{eq:2}
  E(p)=E(0) + \sqrt{p^2+M^2} - M,
\end{equation}
which allows us to extract a physical mass using states at non-zero
spatial momentum. This \emph{kinetic mass} is commonly used to tune
the input quark mass. From figures~\ref{fig:kinups}
and~\ref{fig:kinchm} it is clear that the $b$ quark mass ($M_b=4.0$)
is accurately tuned, while the $c$ quark mass ($M_c=1.0$) is about
$12\%$ too low.

Rearranging expression~(\ref{eq:2}) it is possible to extract the
effective speed of light; \sloppy{$ c^2 =( (\Delta E)^2 + 2M\Delta E
  )/ p^2$}, where $M$ is the kinetic mass determined at some fixed
$\ve{p}$ (taken to be $\ve{p}=1$ in this work).  Of course, this
quantity should be equal to 1, but on the lattice discretisation
effects can ruin this.  We found $c^2=1$ to within statistical errors
of $1\%$ and $4\%$ for the $b\obar{b}$ and $c\obar{c}$ systems
respectively. This compares very favourably with relativistic
formulations without high levels of improvement, where deviations of
up to $10\%$ are not uncommon~\cite{Kayaba:2004hu}.

%\subsection{Systematic Errors}
%\label{sec:systematic-errors}
The systematic way in which NRQCD was developed makes it possible to
estimate the effect of higher order terms left out of the simulation
hamiltonian. To do this the expectation values of these operators are
calculated using potential model estimates of $\<\ve{p}^2\>$ and
$\<\ve{p^4}\>$. These corrections fall naturally into three
categories.

\emph{Relativistic corrections:} The main relativistic correction
comes from the first term left out of the relativistic expansion of
the energy momentum relation\footnote{The correction terms all have an
  extra factor of 2 included because of the two quarks in the
  meson.}, $\<\ve{p}\>^6/8M_Q^5$.

\emph{Radiative corrections:} The effective field theory coefficients
$c_i$ of the terms appearing in the NRQCD hamiltonian take the form
$c_i(M_Q,\alpha_s) = 1 + c^{(1)}_i(M_Q)\alpha_s +\cdots$, where the
$c^{(1)}_i$ are calculable in perturbation theory but are expected to
be $\O(1)$. We estimate the effect of radiative corrections for the
kinetic term; $\alpha_s\<\ve{p}^4\>/4(M_Q)^3$, and for the Darwin
term\footnote{This expression comes from estimating the $\Delta.E$
  term in a potential model.}; $4\pi \alpha_s^2\psi^2(0)^2/(3M_Q)^2$.

\emph{Discretisation corrections:} The terms in the heavy quark
hamiltonian which correct for discretisation errors in the lattice
spatial derivative and temporal derivative have corrections of the
form $\alpha_s a \<\ve{p}^4\>/8n(M_Q)^2$ and $\alpha_s a^2\<p_i^4
\>/12(M_Q)$ respectively.

In this work we are interested only in energy \emph{splittings}, so
that adding errors from these sources for each state in quadrature is
an over estimate. Instead, we calculate the above corrections using
potential model estimates of the \emph{difference} in the expectation
values of the various splittings. The values calculated are shown in
table~\ref{tab:spliterrs}. It can be seen from this table that the
errors are substantially less for the radial splittings than for the
orbital splittings. This is just a statement of the similarity of the
dynamics of the radial states leads to a more efficient cancellation
of higher order effects.

\begin{table}[hc]
  \centering
  \begin{tabular}{c|c|c|c|c}
    Correction & $\Upsilon(2S-1S)$ & $\psi(2S-1S)$ & $\Upsilon(1P-1S)$
    & $\psi(1P-1S)$ \\ \hline
    $\alpha_s\delta p^4 / 4(M_Q)^3$ & $0.55$ & $1.07$ & $2.19$ &
    $1.73$ \\ \hline 
    $\delta p^6 / (M_Q)^5$ & $0.30$ & $1.64$ & $0.92$ & $1.47$  \\ \hline
    $4\pi\alpha_s^2\psi^2(0)/3M^2$ & $0.62$ & $0.90$ & $1.79$ & $2.82$
    \\ \hline 
    \textbf{Total Rel./Rad.} & $\mathbf{0.88}$ & $\mathbf{2.16}$ &
    $\mathbf{2.97}$ & $\mathbf{3.62}$ \\ \hline\hline
    $\alpha_s a^2\delta p_i^4 / 12M_Q$ & $2.2$ & $0.61$ & $8.76$
    & $0.99$  \\ \hline
    $\alpha_s a\delta p^4 / 8M^2n$ & $0.61$ & $0.23$ & $2.45$ & $0.37$
    \\ \hline 
    $4\pi\alpha_s a \psi^2(0)/15$ & $2.47$ & $0.51$ & $7.17$ & $1.60$
    \\ \hline  
    \textbf{Total Disc.} & $\mathbf{3.36}$ & $\mathbf{0.83}$ &
    $\mathbf{11.58}$ & $\mathbf{1.92}$ \\ \hline\hline
    \textbf{Total} & $\mathbf{3.47}$ & $\mathbf{2.31}$ &
    $\mathbf{11.96}$ & $\mathbf{4.09}$ \\ 
  \end{tabular}
  \caption{Systematic corrections to heavy quarkonium splittings as a percentage of the experimental splitting}
  \label{tab:spliterrs}
\end{table}

From these determinations it is clear that the $2S-1S$ splitting of
the $\Upsilon$ system presents the best opportunity for extracting the
lattice spacing for $b\obar{b}$ states, it gives $a^{-1} =
1.121(62)(31)$ where the first error is statistical and the second
comes from the above systematic considerations. Increased statistical
precision means that $\psi(1P-1S)$ splitting the best choice for
setting $a$ from the $c\obar{c}$ system, giving $a^{-1} =
1.111(29)(23)$. The agreement of these two determinations is
unsurprising when using dynamical gauge configurations.

\begin{figure}[ht]
  \begin{minipage}[l]{0.49\linewidth}
    \begin{center}
      \includegraphics[width=1.0\linewidth]{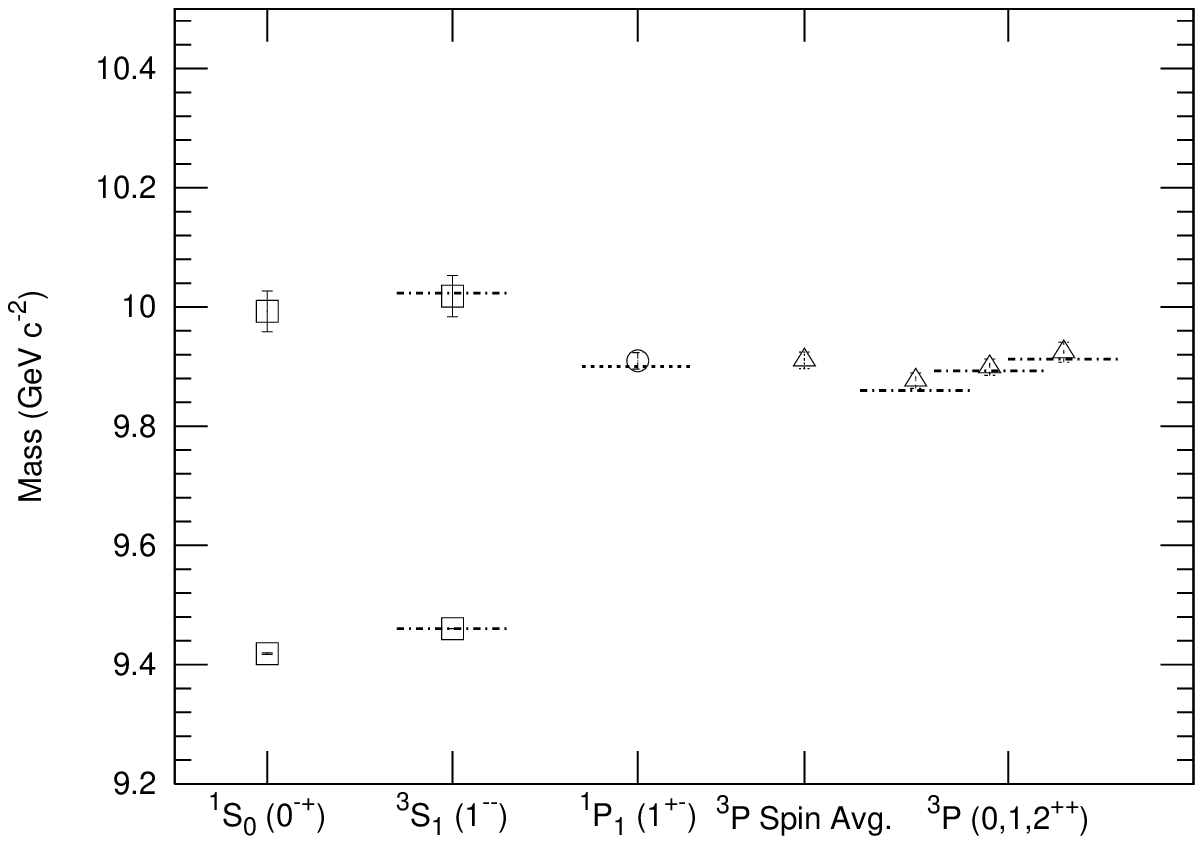}
      \caption{The $\Upsilon$ spectrum. $\Psi (1P-1S)$ has been
        used to set the lattice spacing and the absolute scale has
        been set by the $^3S_1$}
      \label{fig:bbspec}
    \end{center}
  \end{minipage}
  \hspace{2mm}
  \begin{minipage}[r]{0.49\linewidth}
    \begin{center}
      \includegraphics[width=1.0\linewidth]{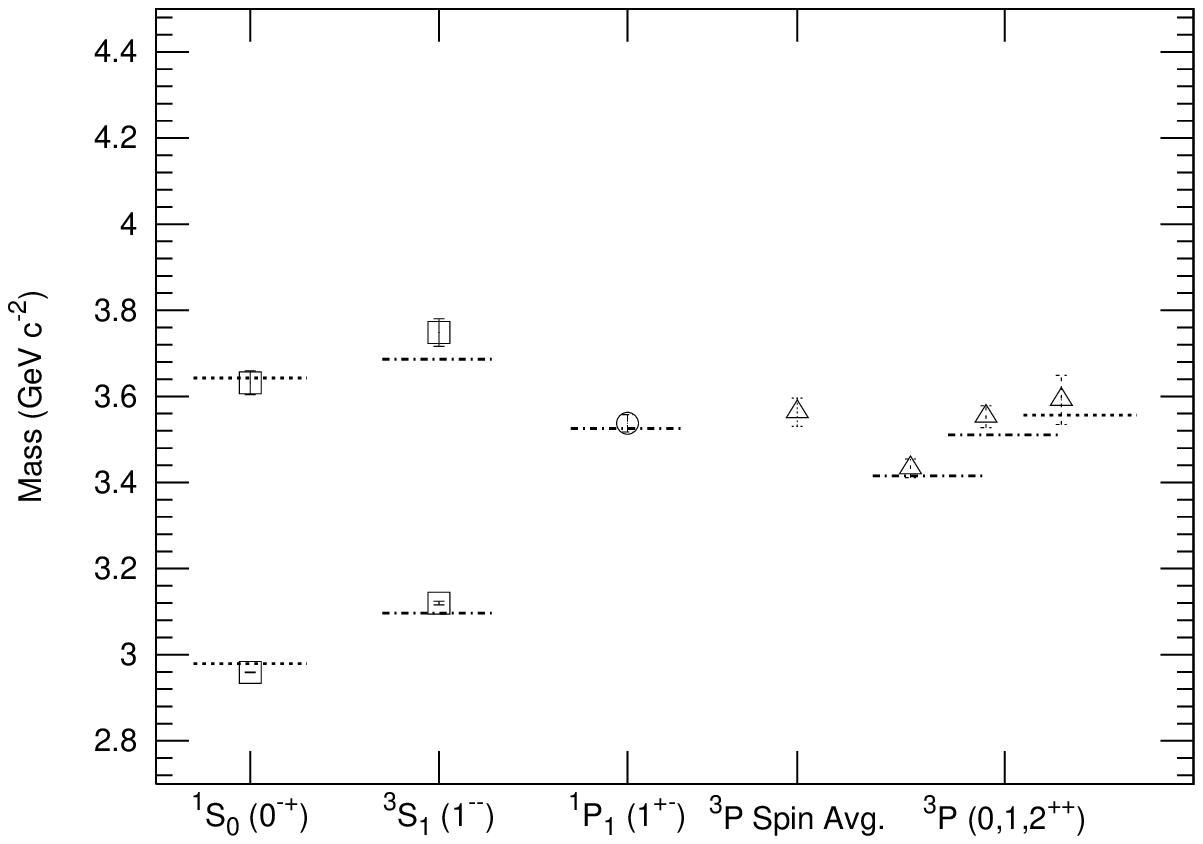}
      \caption{The $\psi$ spectrum. The $\Psi (1P-1S)$ has been used
        to set the lattice spacing and the absolute scale has been
        set at the spin average of the $1S$ states}
      \label{fig:ccspec}
    \end{center}
  \end{minipage}
\end{figure}

In figures~\ref{fig:bbspec} and~~\ref{fig:ccspec} we plot the spectra
of $b\obar{b}$ and $c\obar{c}$ systems respectively.
The agreement of the lattice points with the experimental values
(shown as lines) is good in the $\Upsilon$ system. In the $\psi$
system, there is a noticeable discrepancy in the hyperfine splitting
of $44\mbox{ MeV}$.  This is too large to be fully explained by the
systematic corrections of table~\ref{tab:spliterrs} but is at least in
part due to the fact that the $c$ quark mass was tuned too low.

\section{The $B_c$ Mass}
\label{sec:b_c-mass}
Last year we presented a calculation of the mass of the $B_c$ meson
using NRQCD $b$ quarks and Fermilab method $c$
quarks~\cite{Allison:2004hy}. The use of dynamical configurations in
this calculation allowed a level of precision which, when combined
with the experimental status of the $B_c$ mass at that point, allowed
us to call our calculation a prediction. We found
$M_{B_c}=6.304(4)(11)(^{+18}_{-0})$~\cite{Allison:2004be}, where, in
order, these errors were due to statistics and chiral extrapolation,
uncertainty in setting the input quark masses, and an estimate of
higher order corrections to Fermilab quark action.

Soon after our result was published, new determinations of the $B_c$
mass were made at the Tevatron~\cite{Corcoran:2005ti}. They quoted a
preliminary result of $M_{B_c} = 6.2879(48)(11)$ where the errors were
statistical and systematic respectively.

Our method relies on forming the difference,
\begin{equation}
  \label{eq:5}
  m_{B_c} - 1/2\left(m_\Upsilon+m_{J/\psi\eta_c}\right) = M_{B_c} -
  1/2\left( M_\Upsilon + M_{J/\psi\eta_c}\right)
\end{equation}
where the quantities on the left are lattice energies, and those on
the right are physical masses. The states labelled $\psi\eta_c$ are
the spin average of the $\psi$ and $\eta_c$.  Repeating our earlier
calculation on the extra coarse ensemble but using NRQCD to describe
\emph{both} valence quarks, we find $6.274(1)(25)$, where the first
error is statistical, and the second is calculated in the same way as
those in table~\ref{tab:spliterrs}. Encouragingly, this value is in
agreement with both the experimental value and our previous result.

\section{Heavy light results}
\label{sec:heavy-light-results}
Finally, we also looked at the $B_s$, and $B_d$ systems using and
NRQCD $b$ quark and an improved staggered light quark. This
calculation is complicated by the fact that the fitted correlators
include an oscillatory contribution from the parity partner of the
desired meson. This is a consequence of using staggered light quarks.

\begin{figure}
  \begin{minipage}[l]{0.45\linewidth}
    \begin{center}
      \includegraphics[width=1.0\linewidth]{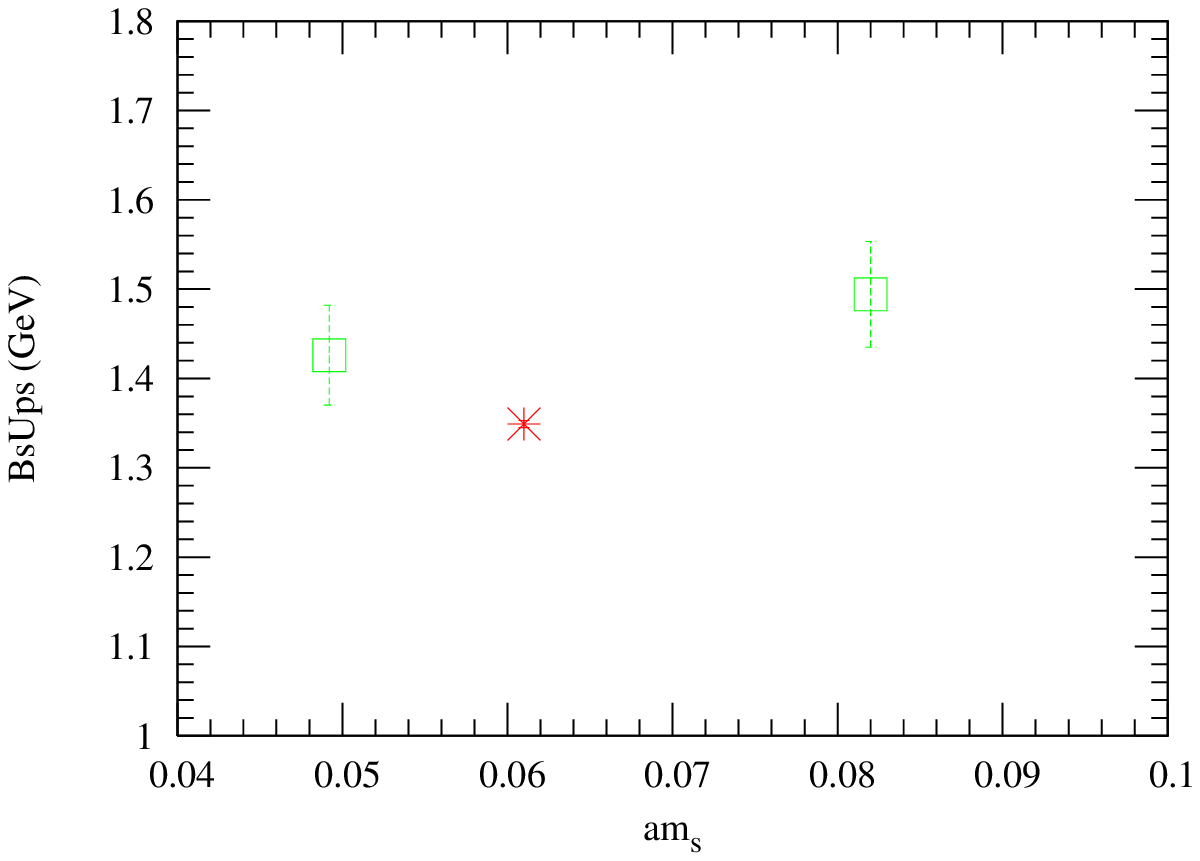}
      \caption{$2m_{B_s}-m_\upsilon$. The boxes are our results and
        the star is experimental value placed at the physical $s$
        quark mass ($am_s=0.062$) for this ensemble.}
      \label{fig:mbs}
    \end{center}
  \end{minipage}
  \hspace{2mm}
  \begin{minipage}[r]{0.45\linewidth}
    \begin{center}
      \includegraphics[width=1.0\linewidth]{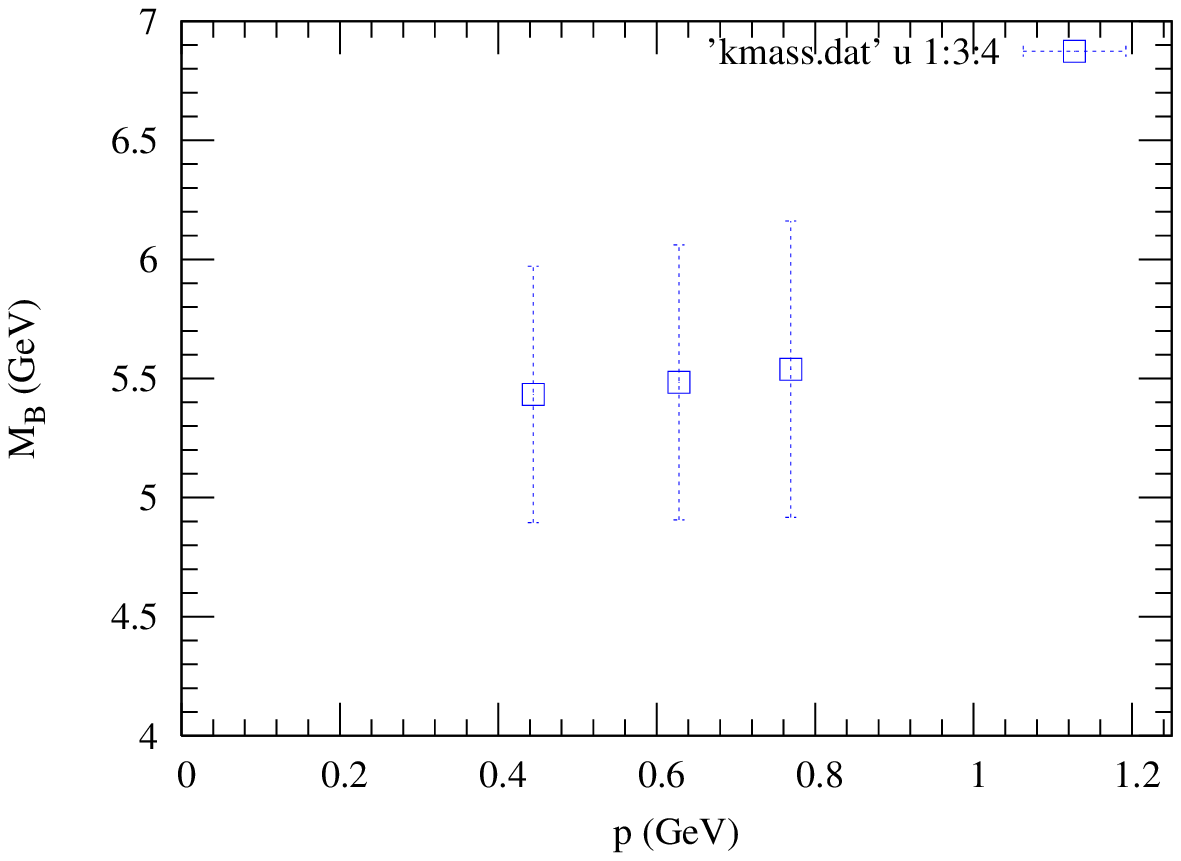}
      \caption{The Kinetic mass of $B_s$ in physical units vs.
        spatial momenta. The errors shown are purely statistical.
        Experimentally $M_{B_s} = 5.3696(24)\mbox{ GeV}$}
      \label{fig:kin-mbs}
    \end{center}
  \end{minipage}
\end{figure}

The results we present here are preliminary, and the reader is
referred to~\cite{Allison:2005ti} for a more complete treatment. In
figure~\ref{fig:mbs} we plot the quantity $2m_{B_s}-m_\Upsilon$. Terms
explicit in the quark mass cancel in this difference, exposing the
differences in the binding energies of the systems.
Our determination of $2m_{B_s}-m_\Upsilon$ seems to be systematically
too large, but we present here only a statistical error and delay a
systematic determination of the errors until~\cite{Allison:2005ti},
where further results will also be presented. It should however be
noted that the sea $s$ quark mass included was estimated to be to high
by approximately $30\%$.

As yet our results for the $B_s$ and $B_d$ do not support any
conclusions, but calculations increasing the number of light quark
masses and simulating the $D_s$ and $D$ with NRQCD $c$ quarks are
currently being analysed. Given the success of our heavy quarkonium
calculations, there is a good prospect that these calculations will
yield useful results.

We thank James Simone and the Fermilab lattice group for the use of
their light quark propagators and computer time at the Fermilab
cluster. Fermilab is operated by the Universities Research Association
Inc. under contract with the US Department of Energy. I.A and C.D are
supported by the UK Particle Physics and Astronomy Research Council.
A.G. was supported by the US Department of Energy.

\bibliographystyle{JHEP-2}
\bibliography{paper}
\end{document}